\documentclass[twocolumn,amssymb,nobibnotes,aps,showpacs,pra]{revtex4}
  \usepackage[dvips]{color}
  \usepackage{newlfont}
  \usepackage{graphicx}
  \usepackage{amssymb}
  \usepackage{amsmath}
  \usepackage[latin1]{inputenc}
  \usepackage{float}
  \usepackage{graphicx}
\newcommand{\br}{\begin{eqnarray}}
\newcommand{\er}{\end{eqnarray}}
\newcommand{\be}{\begin{equation}}
\newcommand{\ee}{\end{equation}}
\newcommand{\nn}{\nonumber}
\def\sig{\mbox{\boldmath $\sigma$}}

\def\hamil{{H}}
\def\rpar{\right)}
\def\lpar{\left(}
\def\rbk{\right]}
\def\lbk{\left[}
\def\rg{\rangle}
\def\lg{\langle}

\begin{document}
\title{Non-Gaussian two-mode squeezing and continuous variable entanglement of linearly and circularly polarized light beams interacting with cold atoms}

\author{R. J. Missori}
\author{M. C. de Oliveira}
\author{K. Furuya}
\affiliation{Instituto de Física ``Gleb Wataghin'', Universidade Estadual de
Campinas,\\ CP 6165, 13083-970, Campinas, SP, Brazil.}
\begin{abstract}
{We investigate how entangled coherent states
and  superpositions of low intensity coherent states of non-Gaussian
nature can be generated via non-resonant interaction between either two linearly or circularly polarized field modes and an ensemble of
X-like four-level atoms placed in an optical cavity. We compare our
results to recent experimental observations and argue that the
non-Gaussian structure of the field states may be present in
those systems.} \pacs{42.50.Dv, 03.67.Mn, 42.50.Ct}
\end{abstract}

\maketitle
\section{Introduction}


Quantum state entanglement is a recognized resource for achieving efficient quantum communication protocols. With the emergence of proposals for continuous variable (CV) quantum communication protocols \cite{BP03,BL05}, such as quantum teleportation \cite{BK98,FU98} and dense coding
\cite{BAN99,BK00,ZP00}, there
has been an increasing interest in the generation and manipulation of entanglement in the CV regime through a diversity of experimental setups.
Recently, bipartite CV entanglement was achieved in a most remarkable experiment, through the interaction of a coherent linearly polarized light beam with a cloud of cold atoms in a high finesse optical cavity \cite{Josse3}. The end product of the interaction is the generation of two entangled squeezed modes with orthogonal polarizations. The entanglement between
those beams is demonstrated by checking the inseparability criterion
for CV \cite{duan,simon}.
 The inseparability criterion is strictly a {\em necessary and sufficient condition} for entanglement only for Gaussian states, but is a {\em sufficient condition} for entanglement for any other CV state. However it is not completely evident whether the bipartite state generated in this experiment is Gaussian or not. Precisely speaking the interaction between the two orthogonally polarized fields intermediated by the atomic cloud is highly expected  to be nonlinear in the field modes annihilation and creation operators. As is well known only Hamiltonians which are at most bilinear in canonically conjugated variables can lead to Gaussian evolution \cite{englert}. A deeper analysis of this system was given in a series of papers  \cite{Josse0,Josse1,Josse2}, and some results were numerically confirmed  \cite{lezama}.
 However
it would be certainly important to stress all the available
possibilities for generating entanglement  and to infer on the Gaussian or non-Gaussian character in this experimental
setup. 

In this paper we investigate a model \cite{Josse0} for the
interaction of two orthogonally polarized quantum fields with an ensemble of $X$-like four level
{atoms}, deriving an effective Hamiltonian
accounting for the field modes interaction. Up to first order the
interaction results to be bilinear in the field operators, which
however is multiplied by the difference of population of an 
effective ensemble of two-level {atoms}, possibly leading thus to a
non-Gaussian evolution. Conditioned on the atomic population
measurement, the two orthogonally polarized fields are left on an
non-Gaussian entangled state, which shows similar properties of the
Gaussian two-mode squeezed vacuum state. Since the experiment allows
for two regimes depending on the detuning between the incident light
and the atomic system, called self-rotation \cite{matsko} or polarization
switching \cite{Josse3}, one deals either with a circularly
polarized beam or a squeezed linearly polarized beam.
 By appropriately
setting one of the input linearly polarized modes in the vacuum
state, we obtain in the output a coherent superposition of coherent
states in the orthogonally polarized linear modes.
 When viewed from the
circular polarization frame, this superposition results in an
entangled coherent state between the two modes in polarization $+$
and $-$. The presence of this superposition, in one polarization
reference frame and an entangled state in the other can explain all
the non-classical features observed in the before mentioned
experiment, with a non-Gaussian state however.  Remarkably recently
much effort has been dedicated to the generation of such
non-Gaussian states in propagating light fields by
photon-subtraction \cite{GR06,POL06}.  In this paper we show that 
in principle such states may be generated in the experiments reported 
in \cite{Josse3,Josse0,Josse1,Josse2} as well.

This paper is organized as follows. In Sec. II we review the
description of the interaction between the two orthogonally
polarized fields with an ensemble of $N$ cold atoms following
\cite{Josse1} and obtain an approximate solution to the
Heisenberg-Langevin equations that govern the atomic { ensemble} 
evolution in the dispersive regime considered. We then derive an effective Hamiltonian accounting for the interaction of an ensemble of two-level atoms and fields in both circular and linear polarizations. We
investigate, in Sec. III, the dynamical generation of entangled
coherent states, and conditional generation of superposition states. In Sec. IV we
analyze the squeezing of quadratures variances for both linear and
circular polarizations and infer on the inseparability of the two
orthogonal modes. Finally, Section V contains a summary and
conclusions.

\section{The Model}
Light fields interacting with cold atoms can show a diversity of interesting 
phenomena such as squeezing, two-mode squeezing and CV entanglement \cite{Josse3,Josse0,Josse1,Josse2}. It can also be employed for implementation of 
quantum logic operations \cite{TUR95}. Recent experiments have demonstrated 
the direct relation between one mode squeezing and two-mode entanglement under
a linear (Bogoliubov) transformation of the fields polarization reference frame
\cite{Josse1,Josse2}. In those  experiments, an $x$-linearly polarized probe 
field is let to interact with a cloud of cold cesium atoms in a high finesse 
optical cavity. The cavity output $x$-polarized signal and the $y$-polarized 
vacuum are squeezed, which results in the entanglement of two orthogonal 
circular polarization fields. The probe light field is red-detuned by about 
50 MHz of the $6S_{1/2},F=4$ to $6P_{3/2},F=5$ transition.
As previously argued \cite{Josse0} this complicated transition can be modeled by an $X$-like four-level atomic structure.

\begin{figure}
\centering
\includegraphics[width=1\columnwidth]{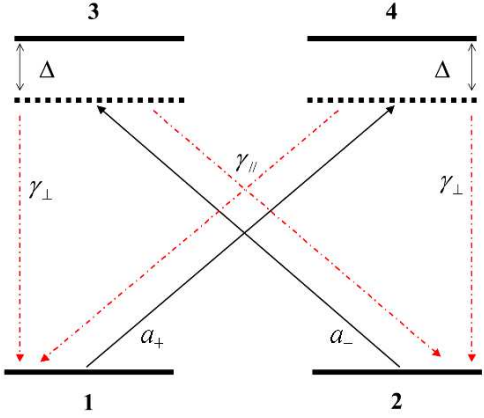}\vspace{-5cm}
\caption{ (Color online) Scheme of the four level X-like atomic model for the $6S_{1/2},F=4$ to $6P_{3/2},F=5$ transition of Cesium.}
\label{fig1}
\end{figure}

Following Ref.\cite{Josse0} we consider the atomic system as being a
set of $N$ $X$-like four-level cold atoms in an optical
cavity driven by a linearly polarized field, 
 as shown schematically in figure \ref{fig1}.
We employ collective operators to describe the N atoms ensemble
(e.g. ${\sigma}_{14}=\sum_{i=1}^{N} e^{i\omega t}|1\rg_{i}\lg 4|_{i}$), and  we denote the operators
related to circularly polarized fields by indexes ${a}_{+(-)}$, which
are defined from the standard linear polarization components
\begin{eqnarray}
{a}_{+}=\frac{{a}_{x}-i {a}_{y}}{\sqrt{2}} \qquad {\mbox{and}}\qquad
{a}_{-}=\frac{{a}_{x}+i {a}_{y}}{\sqrt{2}}\; .
\end{eqnarray}

We shall consider both orthogonally polarized linear modes ${a}_{x}$ and
${a}_{y}$ initially in coherent states  and latter we assume the  mode ${a}_{y}$ in its
vacuum state.
The atomic transition frequencies are chosen in resonance ($\omega_{at}=\omega_{13}=\omega_{24}$) for simplification
matters. In that case, if the field frequency is $\omega$, the
detuning from the atomic {ensemble} transitions resonance
is equal to $\Delta=\omega_{at}-\omega$. The coupling constant
between {the atoms} and field is $g=\epsilon_{0}
d/\hbar$, where $d$ is the atomic dipole and
$\epsilon_{0}=\sqrt{\hbar\omega/2\varepsilon V}$,
 where $V$ is the volume of the cavity. The dipole decay rate $\gamma$, as depicted in
 fig. \ref{fig1}, is decomposed in two orthogonal rates as
 $\gamma=\gamma_{\parallel}+\gamma_{\perp}$. Thus, the {atoms}-field
interaction Hamiltonian is described by  \cite{agarwal}
\begin{eqnarray}
\hamil = \hamil_{0}+\hamil_{I}
\end{eqnarray}
where
\begin{eqnarray}
\hamil_{0}&=&  E_{1}{\sigma}_{11}+E_{2}{\sigma}_{22}+E_{3}{\sigma}_{33}+E_{4}{\sigma}_{44}\nonumber\\
&&+\hbar\omega {a}^{\dag}_{+}{a}_{+}+\hbar\omega {a}^{\dag}_{-}{a}_{-},
\end{eqnarray}
and
\begin{eqnarray}
\hamil_{I} &=& \hbar g\left[e^{-i\omega t}{a}_{+}{\sigma}_{41}+e^{i\omega t}{a}^{\dag}_{+}{\sigma}_{14}\right.\nonumber\\
&& +\left. e^{-i\omega t}{a}_{-}{\sigma}_{32}+e^{i\omega t}{a}^{\dag}_{-}{\sigma}_{23}\right].
\end{eqnarray}
%
The atomic evolution is appropriately governed by a set of quantum Heisenberg-Langevin
equations \cite{Josse0}, here given in a rotating frame with the probe frequency $\omega$
as
\begin{eqnarray}
{\dot{\sigma}_{14}}&=&-(\gamma+i\Delta){\sigma}_{14}-ig{a}_{+}({\sigma}_{11}-{\sigma}_{44})+{F}_{14}, \label{sig1}\\
{\dot{\sigma}_{23}}&=&-(\gamma+i\Delta){\sigma}_{23}-ig{a}_{-}({\sigma}_{22}-{\sigma}_{33})+{F}_{23},\label{sig2}
 \\
{\dot{\sigma}_{11}}&=&
2\gamma_{\perp}{\sigma}_{33}+2\gamma_{\parallel}{\sigma}_{44}-ig({a}^{\dag}_{+}{\sigma}_{14}-h.c.
)+{ F}_{11},\label{sig3}\\
{\dot{\sigma}_{22}}&=&
2\gamma_{\parallel}{\sigma}_{33}+2\gamma_{\perp}{\sigma}_{44}-ig({a}^{\dag}_{-}{\sigma}_{23}-h.c.
)+{ F}_{22},\\
{\dot{\sigma}_{33}}&=&
-2\gamma{\sigma}_{33}+ig({a}^{\dag}_{-}{\sigma}_{23}-h.c.
)+{ F}_{33},
\\
{\dot{\sigma}_{44}}&=&
-2\gamma{\sigma}_{44}+ig({a}^{\dag}_{+}{\sigma}_{14}-h.c.
)+{F}_{44},
\label{sign}
\end{eqnarray}
({ with $\hbar=1$}) where ${ F}_{ij}$ are the Langevin operators.

We want to derive an effective Hamiltonian accounting for the interaction between two- orthogonally polarized field modes. For that we make the following assumptions:

(i) We shall neglect the fluctuation on deriving a stationary solutions to the above set of equations. That means that we shall neglect the Langevin operators by taking ${F}_{ij}=0$.

(ii) To simplify the four level system of atoms to an effective system
of two level atoms, we consider that $\Delta$ is very large ($\Delta\gg\gamma\gg g$) in such a way that the higher levels $3$ and $4$ are not
significantly populated (${\sigma}_{33}={\sigma}_{44}=0$). Thus
by taking Eqs. (\ref{sig1},\ref{sig2}) in the stationary regime ($\dot{\sigma}_{14}=\dot{\sigma}_{23}=0$) results in $(0)$ lowest order approximation
\be
\sigma^{(0)}_{14}=\frac{-ig{a}_{+}(t)\sigma_{11}(t)}{(\gamma+i\Delta)},\label{stationary1}
\ee
and
\be
\sigma^{(0)}_{23}=\frac{-ig{a}_{-}(t)\sigma_{22}(t)}{(\gamma+i\Delta)},
\label{stationary2}
\ee
These solutions can be replaced back into Eqs. (\ref{sig3}-\ref{sign}) for the levels population. Rewriting the interaction term in a symmetrized form we obtain
\br
\dot{\sigma}_{44}&=& \frac{2\gamma g^{2}}{(\gamma^{2}+\Delta^{2})}({a}^{\dag}_{+}{a}_{+}+1/2)\sigma_{11}\nn\\
&& -2\gamma\lbk 1+\frac{g^{2}}{(\gamma^{2}+\Delta^{2})}({a}^{\dag}_{+}{a}_{+}+1/2)\rbk\sigma_{44},\\
\dot{\sigma}_{33}&=&\frac{2\gamma g^{2}}{(\gamma^{2}+\Delta^{2})}({a}^{\dag}_{-}{a}_{-}+1/2)\sigma_{22}\nn\\
&&-2\gamma\lbk 1+\frac{g^{2}}{(\gamma^{2}+\Delta^{2})}({a}^{\dag}_{-}{a}_{-}+1/2)\rbk\sigma_{33},\\
\dot{\sigma}_{22}&=& 2\gamma_{\perp}\sigma_{44}-\frac{2\gamma g^{2}}{(\gamma^{2}+\Delta^{2})}({a}^{\dag}_{-}{a}_{-}+1/2)\sigma_{22} \nn\\
&&+
2\gamma\lbk\frac{\gamma_{\parallel}}{\gamma_{\perp}}+\frac{g^{2}}{(\gamma^{2}+\Delta^{2})}
({a}^{\dag}_{-}{a}_{-}+1/2)\rbk\sigma_{33},\\
\dot{\sigma}_{11}&=& 2\gamma_{\perp}\sigma_{33}-\frac{2\gamma g^{2}}{(\gamma^{2}+\Delta^{2})}({a}^{\dag}_{+}{a}_{+}+1/2)\sigma_{11}\nn\\
&& + 2\gamma\lbk\frac{\gamma_{\parallel}}{\gamma_{\perp}}+\frac{g^{2}}{(\gamma^{2}+\Delta^{2})}
({a}^{\dag}_{+}{a}_{+}+1/2)\rbk\sigma_{44},
\er
where we have already neglected the Langevin terms.
In the stationary regime where the states $3$ and $4$  will be
considered practically not populated 
we find
\br
\sigma_{33}&=&\frac{g^{2}}{(\gamma^{2}+\Delta^{2})}\lbk1+\frac{g^{2}}{(\gamma^{2}+\Delta^{2})}
({a}^{\dag}_{-}{a}_{-}+1/2)\rbk^{-1}\nn\\&&\times({a}^{\dag}_{-}{a}_{-}+1/2)\sigma_{22} ,\label{sig33}\\
\sigma_{44}&=&\frac{g^{2}}{(\gamma^{2}+\Delta^{2})}\lbk1+\frac{g^{2}}{(\gamma^{2}+\Delta^{2})}
({a}^{\dag}_{+}{a}_{+}+1/2)\rbk^{-1}\nn\\&&\times({a}^{\dag}_{+}{a}_{+}+1/2)\sigma_{11} . \label{sig44}
\er
Since we have assumed $\Delta\gg\gamma\gg g$,  we must have $\frac{g^2}{(\gamma^2+\Delta^2)}\ll 1$, and from now on we shall only keep in (\ref{sig33}) and (\ref{sig44}) first order terms in this quantity. That is, we assume \br
\sigma_{33}&\approx&\frac{g^{2}}{(\gamma^{2}+\Delta^{2})})({a}^{\dag}_{-}{a}_{-}+1/2)\sigma_{22} ,\label{sig33n}\\
\sigma_{44}&\approx&\frac{g^{2}}{(\gamma^{2}+\Delta^{2})}({a}^{\dag}_{+}{a}_{+}+1/2)\sigma_{11} . \label{sig44n}
\er 
These stationary solutions for the populations of levels 3 and 4 can be then included in the correspondent first order terms for the coherences between sates 1,4 and 2,3 are as follows
\be
\sigma^{(1)}_{14}=\frac{-ig{a}_{+}(t)\left(\sigma_{11}-\sigma_{44}\right)}{(\gamma+i\Delta)},\label{stationary11}
\ee
and
\be
\sigma^{(1)}_{23}=\frac{-ig{a}_{-}(t)\left(\sigma_{22}-\sigma_{33}\right)}{(\gamma+i\Delta)}.
\label{stationary22}
\ee

We now are in position to consider the Heisenberg equations for the field operators
\br
\dot{{a}}_{+}&=&
-i\omega {a}_{+}-ig{\sig}_{14},\\
\dot{{a}}_{-}&=&
-i\omega {a}_{-}-ig{\sig}_{23},
\label{heisenberg}
\er
where ${\sig}_{14(23)}=e^{i\omega  t}\sigma_{14(23)}$. Substituting the stationary solutions (\ref{stationary11}) and (\ref{stationary22}), we obtain
%
%
 \br
{\dot{{a}}}_{+}&=& -\left\{ i\omega + \frac{ g^{2}}{2(\gamma+i \Delta)}\left[1-\frac{g^{2} a_+a^\dagger_+}{(\gamma^2+\Delta^2)}\right]\right\} {{a}}_{+}\nn\\&& - \frac{ g^{2}}{2(\gamma+i\Delta)} \left[1-\frac{g^{2} a_+a^\dagger_+}{(\gamma^2+\Delta^2)}\right]{a}_{+}{\sig}_{z}\; ,\label{mais1}
\\
{\dot{{a}}}_{-}&=&  -\left\{ i\omega + \frac{ g^{2}}{2(\gamma+i \Delta)}\left[1-\frac{g^{2} a_-a^\dagger_-}{(\gamma^2+\Delta^2)}\right]\right\} {{a}}_{-}\nn\\&& + \frac{ g^{2}}{2(\gamma+i\Delta)} \left[1-\frac{g^{2} a_-a^\dagger_-}{(\gamma^2+\Delta^2)}\right]{a}_{-}{\sig}_{z},
\label{mais2}
\er
where we have defined  ${\sig}_{z}\equiv{\sig}_{11}-{\sig}_{22}$, and we assumed ${\sig}_{11}+{\sig}_{22}\approx \mathbf{1}$. Remark the presence of the non-linear terms proportional to $a_+a_+^\dagger a_+$ and to $a_-a_-^\dagger a_-$. They will lead to non-Gaussian evolutions over the field modes, i.e., if the initial field mode states are Gaussian they will be driven to  non-Gaussian states. 
 The same is true for higher order terms, which imply that invariably the field  operators evolution will be non-Gaussian as inferred in the introduction. 
 These nonlinear terms are at least proportional to $g^4$, which from our assumptions is very small compared to the linear terms in $g^2$, which alone may or not lead to
Gaussian evolutions depending on the prepared atomic system state. Keeping only the terms up to linear in $g^2$, Eqs. (\ref{mais1}) and (\ref{mais2}) simplify to
\br
{\dot{{a}}}_{+}&\approx& -\left[ i\omega + \frac{ g^{2}}{2(\gamma+i \Delta)}\right] {{a}}_{+} - \frac{ g^{2}}{2(\gamma+i\Delta)} {a}_{+}{\sig}_{z}\; ,\label{mais11}
\\
{\dot{{a}}}_{-}&\approx&  -\left[ i\omega + \frac{ g^{2}}{2(\gamma+i \Delta)}\right] {{a}}_{-} + \frac{ g^{2}}{2(\gamma+i\Delta)} {a}_{-}{\sig}_{z},
\label{mais22}
\er
 which, in terms of the linear polarization operators are given by
\br
{\dot{{a}}}_{x}&\approx& -\lbk i\omega + \frac{ g^{2}}{2(\gamma+i \Delta)}\rbk {{a}}_{x} + \frac{i g^{2}}{2(\gamma+i\Delta)} {a}_{y}{\sig}_{z}\; ,
\\
{\dot{{a}}}_{y}&\approx& -\lbk i\omega + \frac{ g^{2}}{2(\gamma+i \Delta)}\rbk {{a}}_{y} - \frac{i g^{2}}{2(\gamma+i\Delta)} {a}_{x}{\sig}_{z} .
\label{heisenberg2}
\er
 Assuming ${\mathbf{a}}_i\equiv e^{\left( i\omega + \frac{ g^{2}}{2(\gamma+i \Delta)}\right) t} {a}_i$, $ i=x,y$ we can rewrite these equations as
 \br
{\dot{{\mathbf a}}}_{x}=\frac{i g^{2}}{2(\gamma+i\Delta)}{{\mathbf a}_{y}}{\sig}_{z} {\mbox{\,\, and \,\, }}{\dot{{\mathbf a}}}_{y}=-\frac{i g^{2}}{2(\gamma+i\Delta)}{{\mathbf a}_{x}}{\sig}_{z}.\label{Heis1}
\er
The appropriate effective interaction Hamiltonians corresponding to the evolution of
the circularly polarized modes corresponding to Eqs. (\ref{mais11},\ref{mais22}) is
\br
{H}_{{eff}}^{(c)}=-i\lambda({\mathbf a}^{\dag}_{+}{\mathbf a}_{+}-{\mathbf a}^{\dag}_{-}{\mathbf a}_{-}){\mathbf {\sig}}_{z},
\label{hcircular}
\er
with
\begin{eqnarray}
\lambda=\frac{g^{2}}{2(\gamma+i\Delta)}&\equiv&\lambda_{1}+ i\;\lambda_{2},
\end{eqnarray}
where \br \lambda_1&\equiv&\frac{g^{2}\gamma}{2(\gamma^{2}+\Delta^{2})},\\
\lambda_2&\equiv&-\frac{\;g^{2}\Delta}{2(\gamma^{2}+\Delta^{2})}.\er

Similarly for the linear polarization with respect to Eqs. (\ref{Heis1}), we obtain
\br
{ H}_{{eff}}^{(l)}=\lambda({\mathbf a}^{\dag}_{y}{\mathbf a}_{x}-{\mathbf a}^{\dag}_{x}{\mathbf a}_{y}){\sig}_{z},
\label{hlinear}
\er
which is a bilinear coupling between the $x$ and $y$ polarization
modes. Had we considered the nonlinear terms as discussed above we
would end up with nonlinear interaction term as well, such as
$a_x^\dagger a_x a_y^\dagger a_y$, which would drive the system
state to a non-Gaussian entangled state such as discussed in
\cite{entmilburn}.

%
%
%
%


Remark that, being $\lambda$ a complex number, both effective Hamiltonians are 
non-Hermitian. The non-Hermiticity comes from the spontaneous emission from 
states 3 and 4 and from the fact that we have eliminated those states in our 
treatment, leading to non-unitary process.
Our choice for the approximations $(\Delta\gg\gamma\gg g)$ are in accordance with
experimental values, which settle $\gamma \approx 2.6 - 16$ MHz, $\Delta\approx 
130-327$ MHz, and $g$ around 2 Hz. To keep such relations between the parameters
 ($\Delta$, $\gamma$, $g$), the appropriate relation for the real and imaginary 
part of the coupling  of the effective Hamiltonians is $|\lambda_{2}|\gg|
\lambda_{1}|$. Thus the effective Hamiltonians assume the following form


%
\br
{ H}_{{eff}}^{(c)}=\lambda_{2}({\mathbf
a}^{\dag}_{+}{\mathbf a}_{+}-{\mathbf a}^{\dag}_{-}{\mathbf
a}_{-}){\mathbf {\sig}}_{z} \label{hcircular1}\er and\br
{ H}_{{eff}}^{(l)}=i\lambda_{2}({\mathbf a}^{\dag}_{y}{\mathbf
a}_{x}-{\mathbf a}^{\dag}_{x}{\mathbf a}_{y}){\sig}_{z},\label{hlinear1}
\er
%

%
%
%

\section{Entangled coherent states and superpositions of coherent states}

 We shall take firstly the Hamiltonians (\ref{hcircular}) and (\ref{hlinear})
in full form and then we consider the limit $|\lambda_1|\ll|\lambda_2|$ to 
properly set the time scale. As we shall see shortly, these Hamiltonians will generate non-Gaussian states only in favorable situations for the population imbalance of the atomic ensemble. { To take into account the possible transitions 
between the two fundamental collective atomic states, we consider one of the two possibilities: 

\textit{(a)} The ensemble of atoms is found in a coherent 
macroscopic superposition of states 1 and 2,  
$|\chi_{a}^{\pm}\rg=(1/\sqrt{2})(|1\rg_{at}\pm|2\rg_{at})$, where $|1(2)\rg_{at} \equiv |11..1(22..2)\rg$.

  \textit{(b)}
Each atom of the ensemble is found in a coherent superposition of states
1 and 2: $ |{\chi^{'}}_{a}\rg= \Pi_{i=1}^N \frac{1}{\sqrt{2}}(|1\rg_i+
|2\rg_i)$. 

Ensembles of atoms prepared in coherent superposition of states interacting with 
light are important for several applications in quantum optics and quantum 
communication \cite{kuz1,kuz2,lukin1,duan2,lukin2,hald,liu} and constitute an essential resource for generation of entanglement for the kind of light field states we analyze.
Although the first situation (a) is not experimentally easy to achieve, it will
illustrate more easily the final non-Gaussian nature of the field states. 
The second situation (b) is possible to be realized by applying non resonant 
classical light $\pi$-pulses to the atomic cloud, initially prepared in one of 
its ground states, before the interaction with the quantum field polarization we 
are considering. Alternatively one could employ more recent techniques to 
generate the coherent superposition, such as the one proposed in Ref. 
\cite{bergmann}.  The field state resulting in this case is a generalization of
the previous one, also clearly non-Gaussian.}

\textit{(i) Circular Polarization.}
  We consider both circularly polarized modes initially prepared in coherent states
($|\alpha\rg_{+}$ and $|\beta\rg_{-}$), where the subscript + or $-$
designate the two orthogonal circular polarizations. Firstly we assume the atomic ensemble state (a)
After some calculation the time
evolved {atoms}-field state is \br |\psi(t)\rg_c &=&
 \frac{1}{\sqrt{2}}[|\alpha e^{\lambda t}\rg_{+}|\beta e^{-\lambda t}
\rg_{-}|1\rg_{at}\nn\\
&&\pm
|\alpha e^{-\lambda t}\rg_{+}|\beta e^{\lambda t}\rg_{-}|2\rg_{at}]\label{circular}
\er
which is obviously in an entangled state. Remark that the bilinear
operation itself do not allow the two fields to be entangled if for
example the atomic ensemble { is found in one of the two states, 
$|1\rg_{at}$ or  $|2\rg_{at}$, or a mixture of them.} The atomic 
superposition of states is an essential resource to generate entanglement.

By conditioning the measurement of the atomic system in the same initial
superposition (a)
, we find
\br
|\varphi(t)\rg_\pm^c &=& \frac{\lg\chi_{a}^{\pm}|\psi(t)\rg}{\sqrt{Tr_{+,-}\{|\lg\chi_{a}^{\pm}|\psi(t)\rg|^2\}}} \nn \\ &=& \frac{1}{{\sqrt{N_\pm}}}
[|\alpha e^{\lambda t}\rg_{+}|\beta e^{-\lambda t}\rg_{-}\pm
|\alpha e^{-\lambda t}\rg_{+}|\beta e^{\lambda t}\rg_{-}],\nn\\
\label{ecs1}
\er where
\br N_\pm&=&2\left\{1\pm e^{-(|\alpha|^2+|\beta|^2)\cosh{2\lambda_1t}}e^{(|\alpha|^2+|\beta|^2)\cos{2\lambda_2t}}\right.\nn\\&&\times\left.\cos[(|\alpha|^2-|\beta|^2)\sin 2\lambda_2t]\right\} \label{norm}
\er
Since $\lambda$ has both real and imaginary parts we see that as the time 
evolves the two modes become more and more entangled, due both to an increase 
of the coherent states amplitudes and to a time dependent dephasing 
between the two components of the superposition. To illustrate 
we set the time scale for
$|\lambda_{2}|t=\pi/2$, obtaining 
\br
|\varphi(\pi/2|\lambda_{2}|)\rg_\pm^c &=&
\frac{1}{\sqrt{N_{\pm}}}[|e^{\frac{\pi\lambda_{1}}{2\lambda_{2}}}(i\alpha)\rg_{+}
|e^{-\frac{\pi\lambda_{1}}{2\lambda_{2}}}(-i\beta)\rg_{-} \nn\\&&
\pm |e^{-\frac{\pi\lambda_{1}}{2\lambda_{2}}}(-i\alpha)\rg_{+}
|e^{\frac{\pi\lambda_{1}}{2\lambda_{2}}}(i\beta)\rg_{-}].\label{ecs12}
\er
%
Taking the $\lambda_{1}/|\lambda_{2}|
\rightarrow0$ limit in Eq. (\ref{ecs12}), $|\varphi(\pi/2|\lambda_{2}|)\rg_\pm^c$ assumes the following form
\br
 \frac{1}{\sqrt{N_{\pm}}}[|(i\alpha)\rg_{+}
|(-i\beta)\rg_{-} \pm |(-i\alpha)\rg_{+} |(i\beta)\rg_{-}],
\label{ecs11}
\er
which is an entangled, but {\em non-Gaussian} state.

{ To show that such non-Gaussian entangled state would be present in many experimental configurations let us consider the much simpler to achieve state (b). Following the same procedure, but for the second initial atomic ensemble 
state (b), 
 we obtain after the conditioned measurement of the atomic system in the 
same initial ensemble of superposition states, the following state
\br
|\varphi^{'}(t)\rg^c &&= \frac{\lg {\chi^{'}}_{a}|\psi(t)\rg}{\sqrt{Tr_{+,-}\{|\lg{\chi^{'}}_{a}|\psi(t)\rg|^2\}}} = \nn \\ && \frac{1}{{\sqrt{N^{'}(t)}}}
\left[ \sum_{j=0}^{N} C_j^N | \alpha e^{\frac{N-2j}{N}\lambda t}\rg_{+}|\beta  e^{-\frac{N-2j}{N}\lambda t}\rg_{-} \right],\nn\\
\label{ecs1'}
\er
where $C_j^N= \frac{N!}{j!(N-j)!}$ is the binomial coefficient, and $N^{'}(t)$ 
the corresponding  normalization factor. This is a two mode non-Gaussian entangled 
superposition state with $N+1$ terms in the superposition. Despite being simpler to be experimentally realized this last state has a cumbersome structure than the previous case Eq. (\ref{ecs1}). For illustration we want to keep the simplest non-Gaussian state, and thus from now on we consider only the results for the coherent superposition state (a), but the results 
should follow in similar fashion for the experimentally more accessible situation (b). }

\textit{(ii) Linear Polarization.}
  Under the same initial conditions
(both field modes in the circular polarization prepared in coherent states and the
atoms in state (a), we consider the effect of the evolution operator
\begin{eqnarray}
{U}^{(l)}(t)=\exp{\lbk -i\lambda({\mathbf a}^{\dag}_{y}{\mathbf a}_{x}-{\mathbf a}^{\dag}_{x}{\mathbf a}_{y}){\sig}_{z}t\rbk}.\label{ev}
\end{eqnarray} 
for the linear polarization.
This is a beam-splitter operation with a phase conditioning on the atomic state. It is well
known \cite{knight,bin,Marcos1noncl} that the bilinear beam-splitter operation do not
entangle classical states. As a consequence, if the states of mode ($x$) and ($y$) are
prepared in coherent states they evolve as coherent non-entangled states, unless the
atomic system is prepared in a superposition of the two ground states, 1 and 2.

The two polarization frames are related by the field operators from Eq. (1).
By employing this relation it is immediate that the two initial states are related by
\be |\alpha\rg_+ |\beta\rg_-\rightarrow |\alpha^\prime\rg_y |\beta^\prime\rg_x,\ee
where $\alpha^\prime\equiv i{\left(\alpha-\beta\right)}/{\sqrt 2}$ and
$\beta^\prime\equiv {\left(\alpha+\beta\right)}/{\sqrt 2}$.
Under the evolution (\ref{ev}) the above state takes the form
\begin{widetext}
\br
|\psi(t)\rg_{\pm}^l &=& \frac{1}{\sqrt{2}}{ U}^{(l)}(t){D}_{y}(\alpha^\prime){ D}_{x}(\beta^\prime)|0\rg_{y}|0\rg_{x}\lpar |1\rg_{at}\pm |2\rg_{at}\rpar
 \nn\\
&=& \frac{1}{\sqrt{2}} 
{D}_{y}\lpar \cosh{(\lambda t)}\alpha^\prime-i\sinh{(\lambda t)}\beta^\prime\rpar { D}_{x}\lpar \cosh{(\lambda t)}\beta^\prime+i\sinh{(\lambda t)}\alpha^\prime\rpar|0\rg_{y}|0\rg_{x}|1\rg_{at} \nn\\& & \pm \frac{1}{\sqrt{2}}{ D}_{y}\lpar \cosh{(\lambda t)}\alpha^\prime+i\sinh{(\lambda t)}\beta^\prime\rpar { D}_{x}\lpar \cosh{(\lambda t)}\beta^\prime-i\sinh{(\lambda t)}\alpha^\prime\rpar|0\rg_{y}|0\rg_{x}|2\rg_{at}\nn\\
&=& \frac{1}{\sqrt{2}} |\cosh{(\lambda t)}\alpha^\prime-i\sinh{(\lambda t)}\beta^\prime \rg_{y}\otimes
|\cosh{(\lambda t)}\beta^\prime+i\sinh{(\lambda t)}\alpha^\prime\rg_{x}\;\;|1\rg_{at} \nn \\
&&\pm \frac{1}{\sqrt{2}} |\cosh{(\lambda t)}\alpha^\prime+i\sinh{(\lambda t)}\beta^\prime \rg_{y}\otimes
|\cosh{(\lambda t)}\beta^\prime-i\sinh{(\lambda t)}\alpha^\prime\rg_{x}\;\;|2\rg_{at},
\er 
\end{widetext}
where $D_i(\alpha)$ is the displacement operator in $\alpha$ respective to the polarization $i=x,y$.
Conditioning the atomic detection into the same superposition state (a) prepared at the beginning,
we have
\begin{widetext}
\br
|\varphi(t)\rg_{\pm}^l = \frac{\lg\chi_{a}^{\pm}|\psi(t)\rg}{\sqrt{Tr_{+,-}\{|\lg\chi_{a}^{\pm}|\psi(t)\rg|^2\}}}
&=& \frac{1}{\sqrt{N_{\pm}}}[|\cosh{(\lambda t)}\alpha^\prime-i\sinh{(\lambda t)}\beta^\prime\rg_{y}
\otimes |\cosh{(\lambda t)}\beta^\prime+i\sinh{(\lambda t)}\alpha^\prime\rg_{x} \nn \\ &&
\pm |\cosh{(\lambda t)}\alpha^\prime+i\sinh{(\lambda t)}\beta^\prime\rg_{y}\nn
\otimes |\cosh{(\lambda t)}\beta^\prime-i\sinh{(\lambda t)}\alpha^\prime\rg_{x}]. 
\label{plin1}
\er
Again for the choice $|\lambda_{2}|t=\pi/2$, we obtain 
\br
|\varphi(\pi/2|\lambda_{2}|)\rg_\pm^l &=&
 \frac{1}{\sqrt{N_{\pm}}}
\bigg|i\sinh{(\frac{\pi\lambda_{1}}{2\lambda_{2}})}\alpha^\prime+\cosh{(\frac{\pi\lambda_{1}}{2\lambda_{2}})}\beta^\prime\bigg\rg_{y}
\otimes \bigg|i\sinh{(\frac{\pi\lambda_{1}}{2\lambda_{2}})}\beta^\prime-\cosh{(\frac{\pi\lambda_{1}}{2\lambda_{2}})}\alpha^\prime\bigg\rg_{x} \nn \\ &&
\pm\frac{1}{\sqrt{N_{\pm}}}
\bigg|i\sinh{(\frac{\pi\lambda_{1}}{2\lambda_{2}})}\alpha^\prime-\cosh{(\frac{\pi\lambda_{1}}{2\lambda_{2}})}\beta^\prime\bigg\rg_{y}
\otimes\bigg|i\sinh{(\frac{\pi\lambda_{1}}{2\lambda_{2}})}\beta^\prime+\cosh{(\frac{\pi\lambda_{1}}{2\lambda_{2}})}\alpha^\prime\bigg\rg_{x},
\label{plin2}
\er
\end{widetext}
which for $\lambda_{1}/|\lambda_{2}|\rightarrow 0$ turns out to be
%
\br
|\varphi(\pi/2|\lambda_{2}|)\rg_\pm^l &=& \frac{1}{\sqrt{N_{\pm}}}
\left[\left|\frac{\left(\alpha+\beta\right)}{\sqrt 2}\right\rg_{y}\left|-i\frac{\left(\alpha-\beta\right)}{\sqrt 2}\right\rg_{x}\right.\nn\\&&\left. \pm \left|-\frac{\left(\alpha+\beta\right)}{\sqrt 2}\right\rg_{y}\left|i\frac{\left(\alpha-\beta\right)}{\sqrt 2}\right\rg_{x}\right].
\label{plin3}
\er

To match further with the experimental conditions we set the $x$-polarized mode in a
vacuum state, such that $\alpha-\beta=0$, obtaining
\br
|\varphi(\pi/2|\lambda_{2}|)\rg_\pm^l&=& \frac{1}{\sqrt{N_{\pm}}}\lpar \left|\sqrt{2}\alpha\right\rg_{y} \pm \left|-{\sqrt 2}\alpha\right\rg_{y}\rpar|0\rg_{x}.\nn\\
\label{plin4}
\er
This state represents a superposition of coherent states for the $y$-polarized mode,
namely the odd or even coherent state \cite{dodonov} for the + or - choice for the
atomic state. This state is of a remarkable importance for many applications on quantum
information and computation \cite{marcosmunro}. Now, in the circular polarization Eq. (\ref{ecs11}), the state follows
\br
|\varphi(\pi/2|\lambda_{2}|)\rg_\pm^c &\approx&\frac{1}{\sqrt{N_{\pm}}}[|(i\alpha)\rg_{+}
|(-i\alpha)\rg_{-} \nn\\&&\pm
|(-i\alpha)\rg_{+}
|(i\alpha)\rg_{-}],
\label{ecs111}
\er
and unlike the linear polarization case (superposition in one mode and vacuum in the
orthogonal mode), the two orthogonal circular polarization modes are entangled. This is a particular realization of quasi-Bell states, the so-called entangled coherent states,as fully discussed in \cite{sanders}, and
employed in \cite{entmilburn}.


To end this section we remark that through our calculations we have neglected dissipative
effects over the two field modes. Obviously dissipative effects will be always present
and will drive the system to mixed states, reducing the amount of entanglement. However,
the inclusion of dissipative effects would not alter the non-Gaussian nature
of the field modes. Thus, since we want to keep the evolution as close as possible from
a Gaussian one, and to simplify our discussion, we will not consider the effects of
dissipation for the field modes.

\section{Quadrature Variances and Entanglement Criteria}
Entanglement in the circular polarization can be
inferred from non-classicality signatures  on the linear polarization (see e.g. \cite{Marcos1noncl,marcos1}). This is commonly
realized by analyzing squeezing of the quadratures variances, as in Ref. \cite{Josse1}. Employing the
usual definition of the quadrature operators
\br{X}_{l}=\frac{1}{2}(\mathbf{a}_{l}+\mathbf{a}_{l}^{\dag})\qquad
{Y}_{l}=\frac{1}{2i}(\mathbf{a}^\dagger_{l}-\mathbf{a}_{l}),
\er
where $l=+,-,x,$ or $y$ for each of the circular or linear polarization mode, the variance for the  $X_l$ quadrature is given by $
(\Delta X_{l})^{2}=\lg {X}_{l}^{2} \rg - \lg {X}_{l}\rg ^{2},
$ and similarly for $Y_l$.
These two quadratures in a polarization frame can be combined to indicate if there is entanglement in another polarization frame through the CV inseparability criterion \cite{duan,simon}. In terms of variance operators
 it is given by
\br
I_{a,b}=\frac{1}{2}[\Delta^{2}(X_{a}+X_{b})+\Delta^{2}(Y_{a}-Y_{b})]<\; 1.
\er
For Gaussian states, $I_{a,b}<1$ is a necessary and sufficient condition for entanglement.
In our case where the states are non-Gaussian, while $I_{a,b}<1$ undoubtedly indicates
entanglement, when $I_{a,b}\ge1$ nothing can be said about the presence of entanglement. In this section we analyze the behavior of the quadratures variances as indicators of entanglement and the inseparability criteria for the non-Gaussian states obtained. These results should be compared to the ones of Fig. 2 from Ref. \cite{Josse1}, attributed to Gaussian states.

\textit{(i) Quadrature variances for the circularly polarized modes.}  The quadrature variances calculated
%
from the state of Eq. (\ref{ecs1}),
%
and with parameters fixed to reproduce the experimental situation \cite{Josse0} in the case the $x$-polarized mode is in a vacuum state ($\alpha=\beta$) are depicted in Fig 2.
\begin{figure}[!ht]\begin{center}\vspace{-1cm}
\includegraphics[scale=0.4]{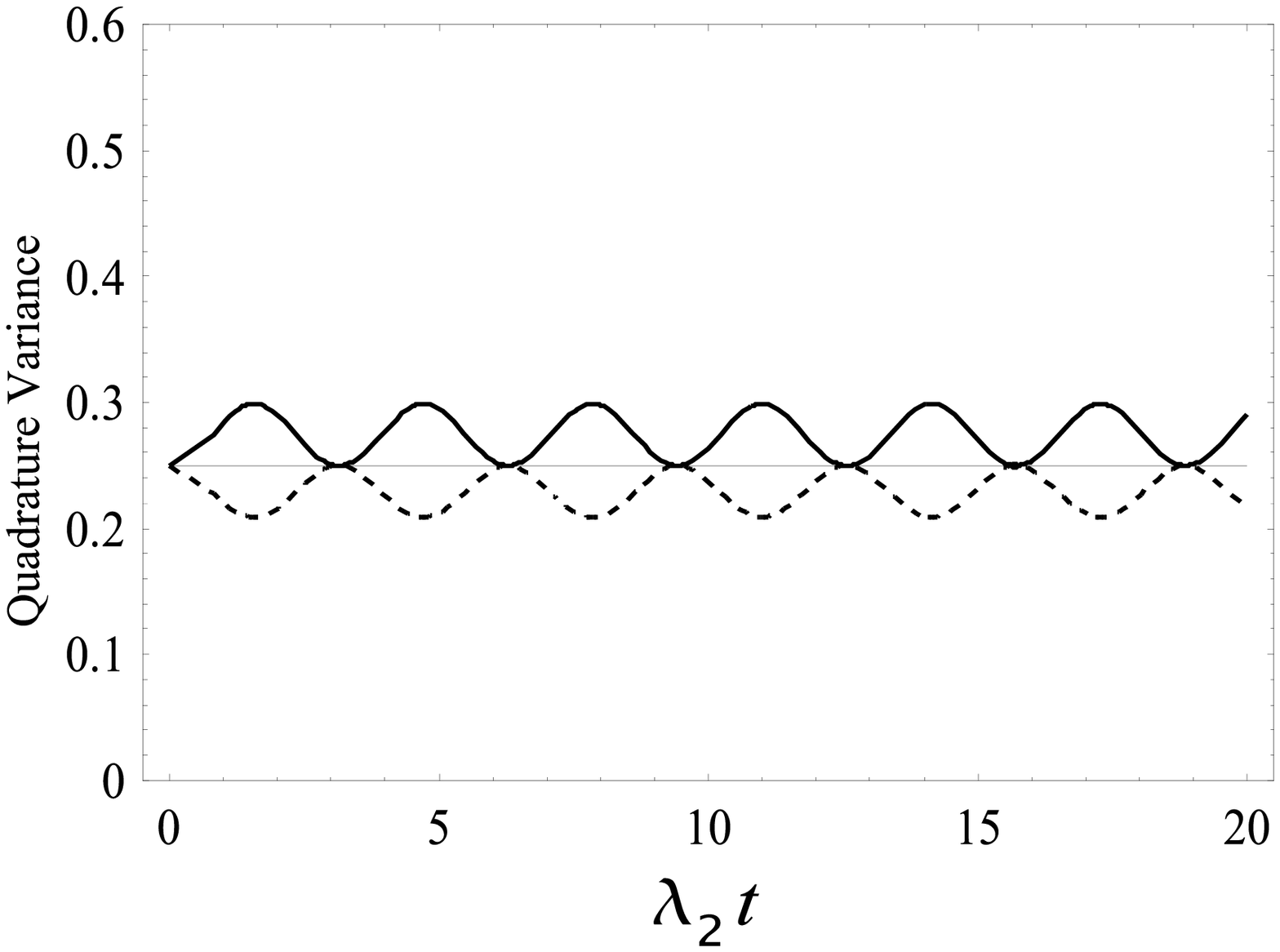}

\vspace{-5cm}\includegraphics[scale=0.4]{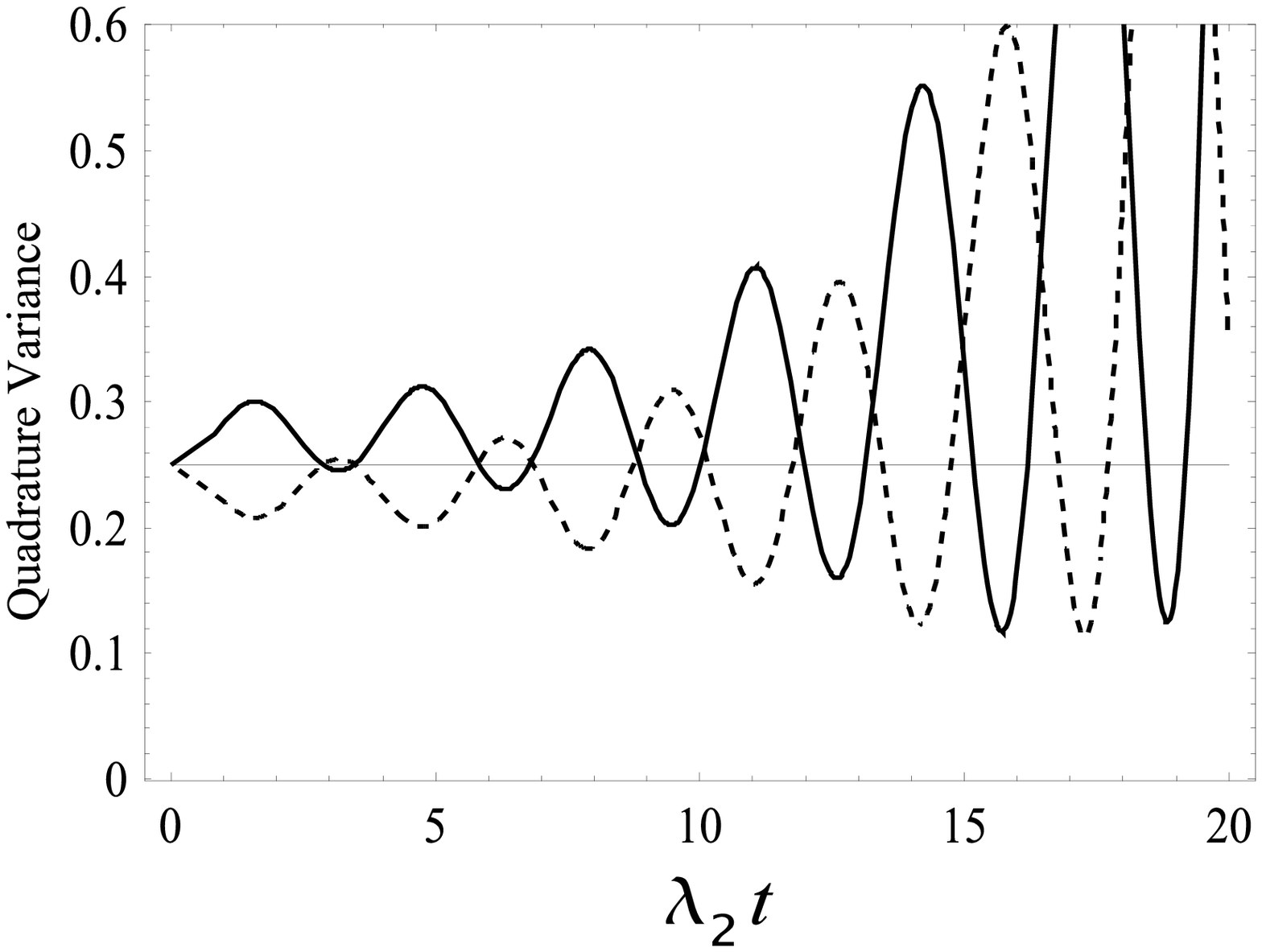}\end{center}
\vspace{-4cm}
\caption{{ Time evolution of the} quadrature variances for circularly polarized modes, with $\beta=\alpha$, $Re[\alpha]=0$ and $Im[\alpha]=0.3$. (a) $(\Delta X_{+})^{2}$ (continuous line) and $(\Delta Y_{+})^{2}$ (dashed line) for $\lambda_{1}/|\lambda_{2}|=0$. (b) $(\Delta X_{+})^{2}$ (continuous line) and $(\Delta Y_{+})^{2}$ (dashed line) for $\lambda_{1}/|\lambda_{2}|=0.1$.}
\label{fig2}
\end{figure}
  In Fig.\ref{fig2}(a) we see an oscillatory (periodic) behavior for both quadratures
  variances of the + circularly polarized  mode. Notice that the variance for the
  quadrature $Y_+$ is periodically compressed, oscillating bellow the reference
  line at the value $1/4$, while the variance for $X_+$ always oscillates above
  this line. In Fig.\ref{fig2}(b), the only modification relative to the previous
  case is the value of the ratio between the $\lambda$'s ($\lambda_{1}/|\lambda_{2}|=0.1$)
  In that case the variance of the quadrature $X_+$ is compressed as well with time. We see
  a similar behavior, but for longer times the amplitude of the oscillations increases due
  to the non-Hermitian nature of the Hamiltonian. The variances $X_-$ and $Y_-$ show exactly
  the same behavior of the ones for the + polarization.

\begin{figure}[!ht]
\includegraphics[scale=0.4]{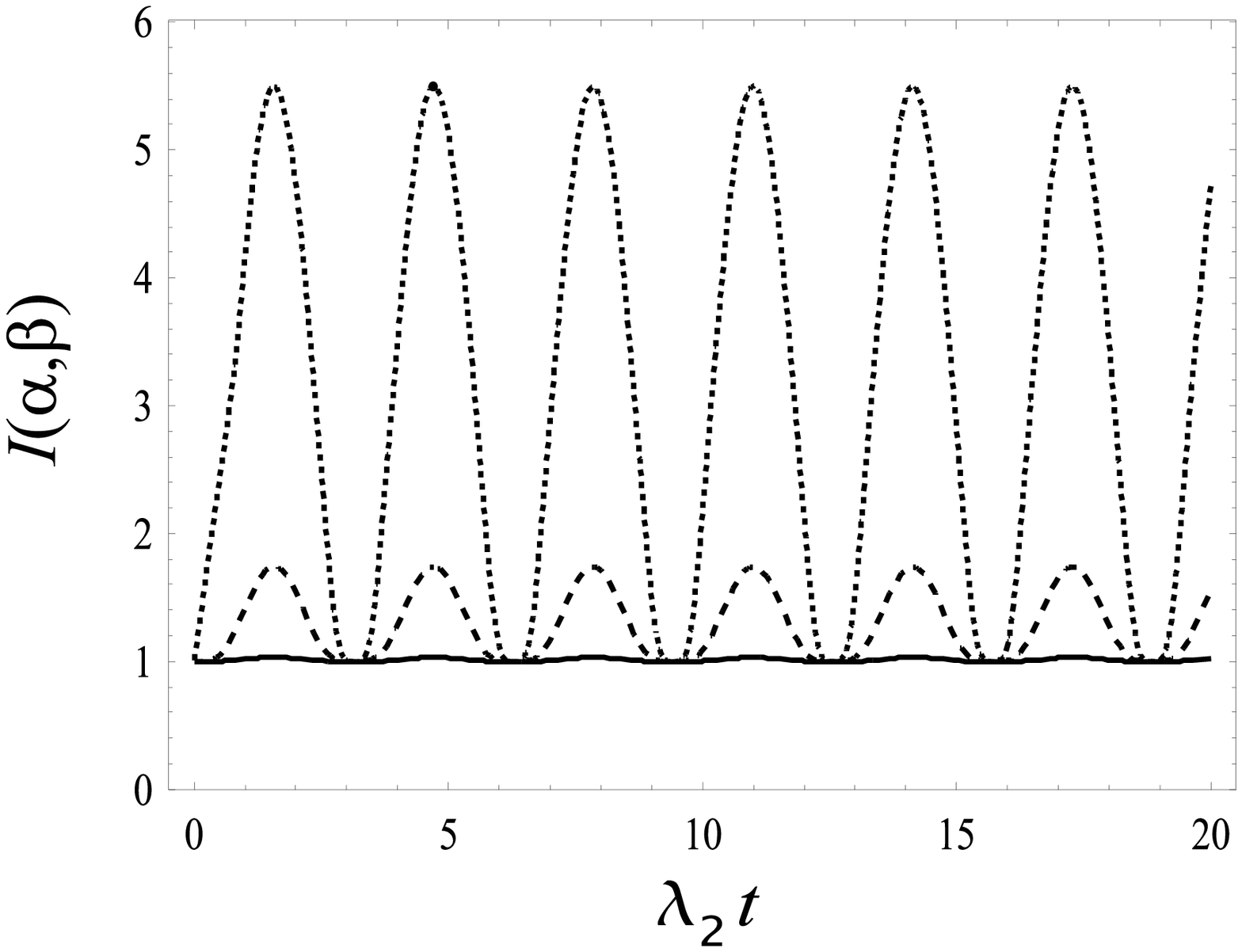}

\vspace{-4cm}
\includegraphics[scale=0.39]{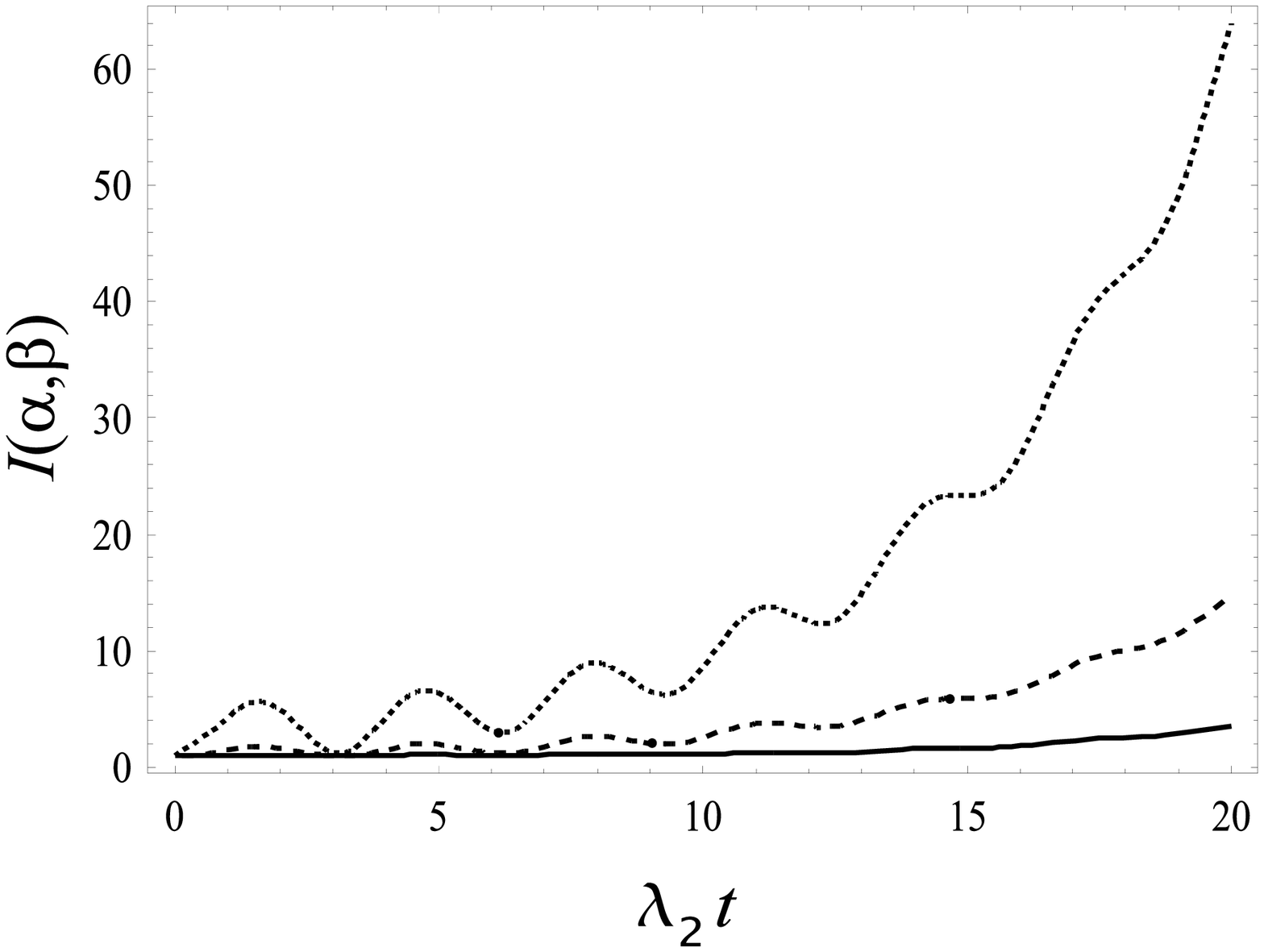}\vspace{-4cm}
\caption{Plot of the { time evolution of the} inseparability criteria for the linear polarization for $\alpha=\beta$ fixing three different lines for $Re{[\alpha]}=0$, $Im{[\alpha]}=0.3$ (continuous line), $Im{[\alpha]}=0.7$ (dashed line), and  $Im{[\alpha]}=1.5$ (short-dashed line). (a)  $\lambda_{1}/|\lambda_{2}|=0$ and (b) $\lambda_{1}/|\lambda_{2}|=0.1$. All plots indicate no entanglement in accordance with the separable, but not Gaussian, state obtained in this situation.}
\label{fig3}
\end{figure}

For the linearly polarized modes the criteria writes
\br
I_{x,y}(\alpha,\beta)&=&\frac{1}{2}[\Delta^{2}(X_{x}+X_{y})(\alpha,\beta)\nn\\&&+\Delta^{2}(Y_{x}-Y_{y})(\alpha,\beta)]<\; 1,
\er
and for the situation $\alpha=\beta$ the inequality correctly indicates no entanglement, since the $x$ mode is left in a vacuum state. The plots for this case are shown
in Fig.\ref{fig3}. Fig.\ref{fig3}(a) is for $\lambda_{1}/|\lambda_{2}|=0$, while fig. 3(b)
is for $\lambda_{1}/|\lambda_{2}|=0.1$. In all situations the inequality is violated, which
in this case correctly indicates no entanglement.

\begin{figure}[!ht]\begin{center}\vspace{-1cm}
\includegraphics[scale=0.4]{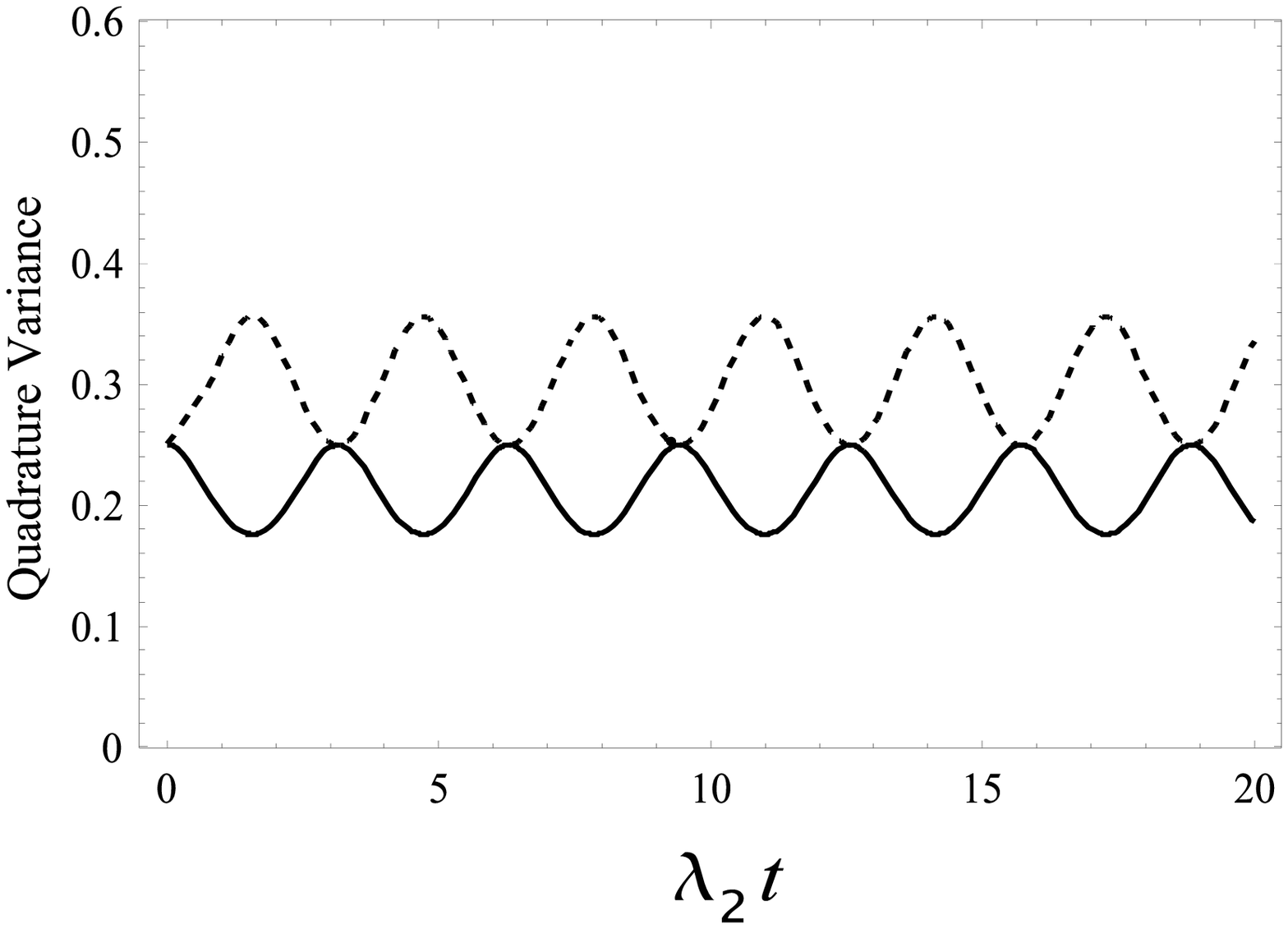}

\vspace{-5cm}
\includegraphics[scale=0.4]{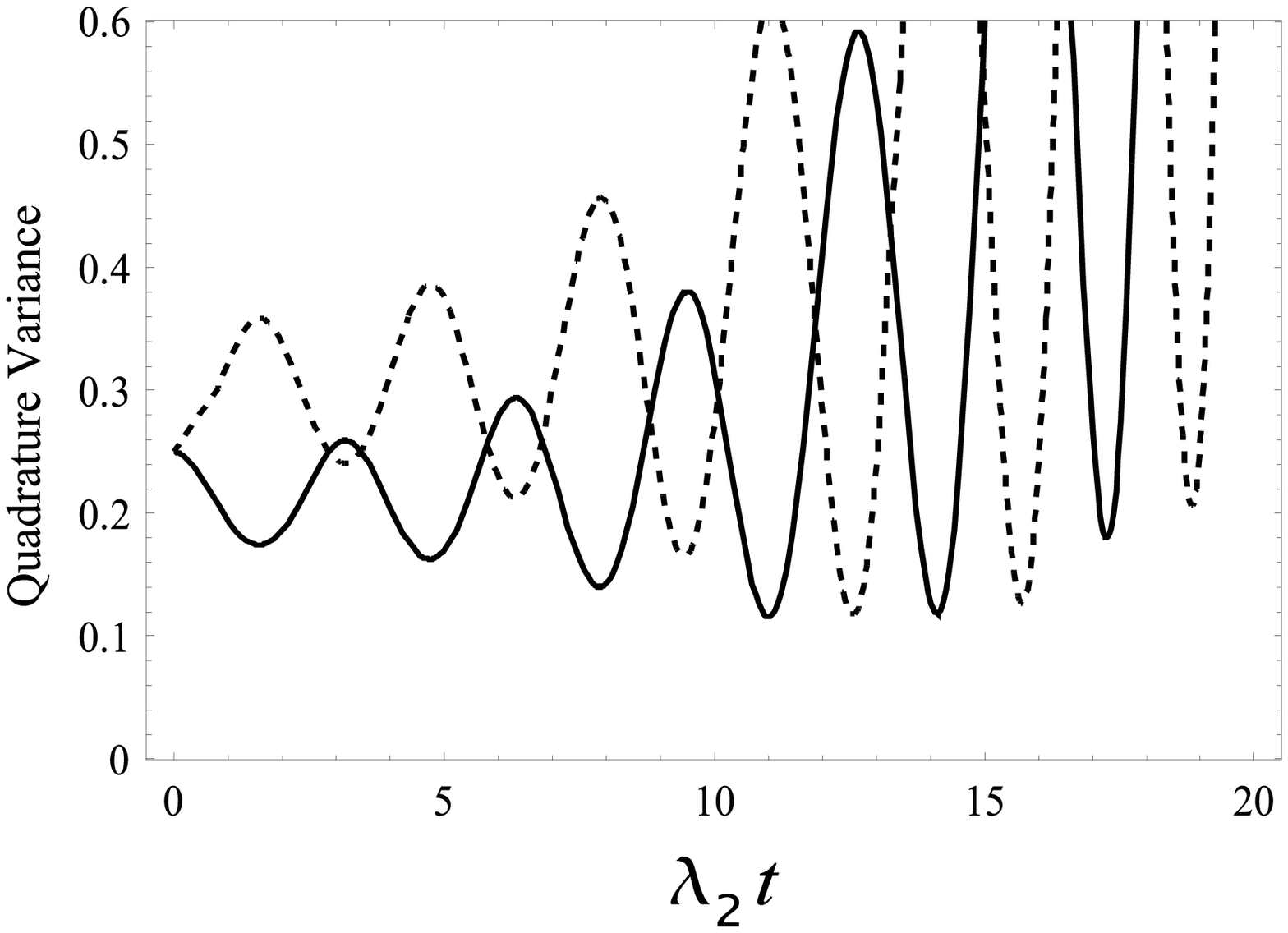}\vspace{-4cm}
\end{center}
\caption{{ Time evolution of} quadrature variances for linearly polarized modes, with $\alpha=\beta, Re[\alpha]=0$ and $Im[\alpha]=0.3$. (a) $(\Delta X_{y})^{2}$ (solid line) and $(\Delta Y_{y})^{2}$ (dashed line) for $\lambda_{1}/|\lambda_{2}|=0$. (b) $(\Delta X_{y})^{2}$ (solid line) and $(\Delta Y_{y})^{2}$ (dashed line) for $\lambda_{1}/|\lambda_{2}|=0.1$. Mode $x$ is left in a vacuum state.}
\label{fig4}
\end{figure}

\textit{(ii) Quadrature variances for the linearly polarized modes.}
For the linearly polarized modes we shall consider the variances from the
%
state of Eq. (\ref{ecs1})
%
 with  $\alpha=\beta$, and $\lambda_{1}/|\lambda_{2}|\rightarrow 0$. As can be noted in the plots of Fig.\ref{fig4}(a) for this case
 and $\lambda_{1}/|\lambda_{2}|=0$, only the variance of the $X_y$ quadrature is
 squeezed (always below the reference line), while the $x$-mode is left in a vacuum state. Again in Fig. 4(b) we
 take $\lambda_{1}/|\lambda_{2}|=0.1$ and remark that the variance of the quadrature $Y_y$
 is squeezed as well with time as an effect of the non-Hermiticity.

\begin{figure}[!ht]
\includegraphics[scale=0.4]{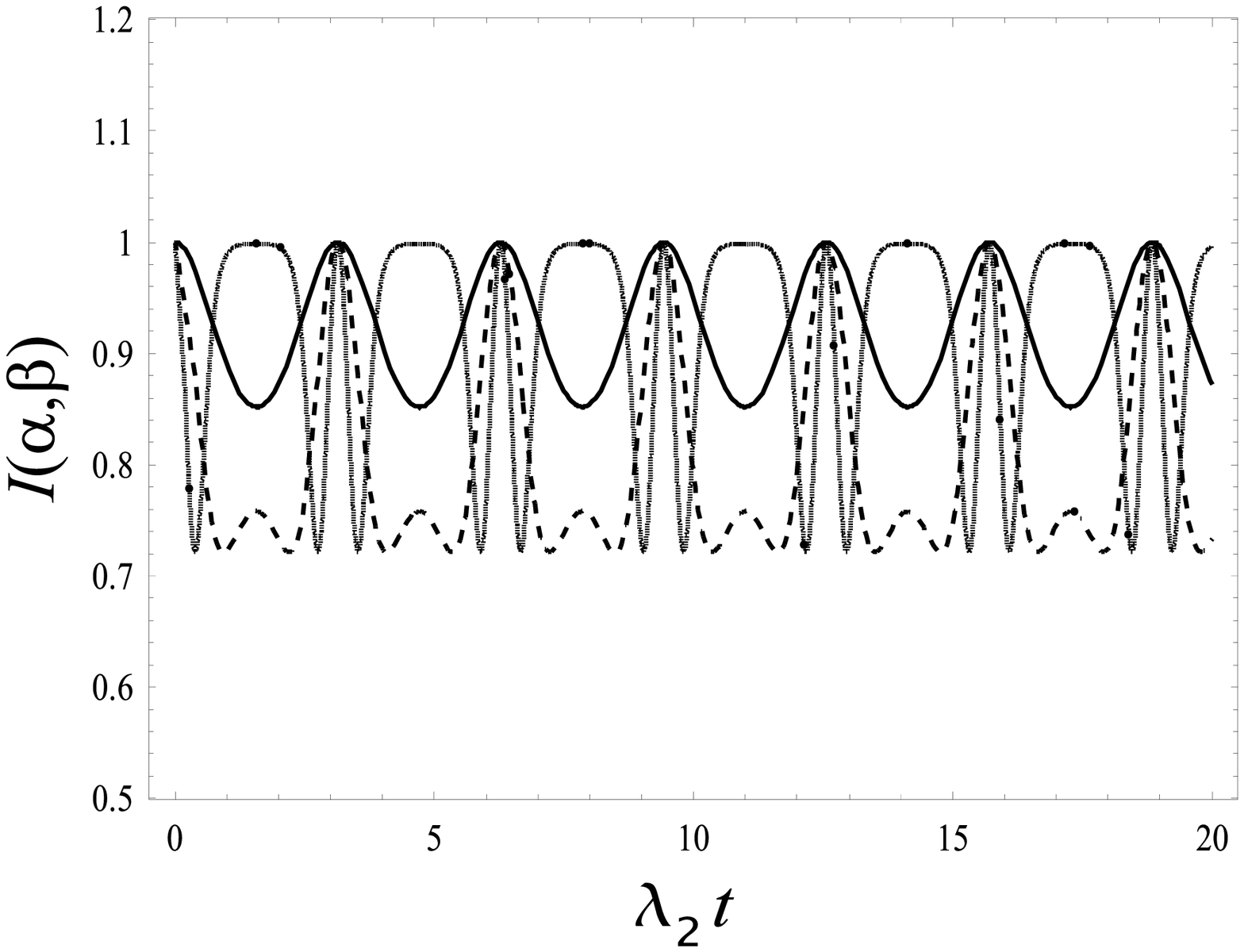}

\vspace{-4cm}
\includegraphics[scale=0.4]{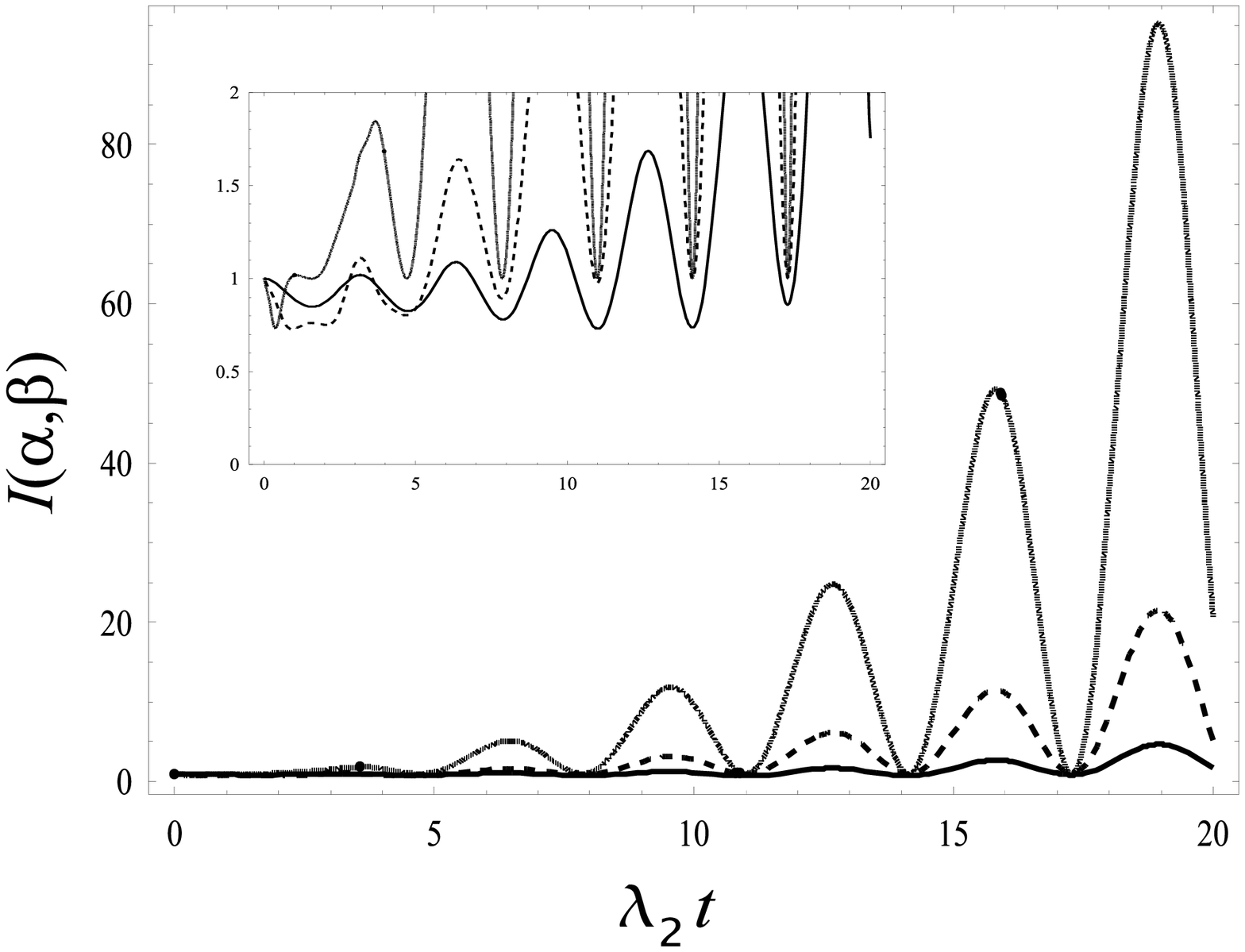}\vspace{-4cm}
\caption{Plot of the inseparability criteria { as a function of time} for the circular polarization modes for $\alpha=\beta$ fixing three different lines for $Re{[\alpha]}=0$, $Im{[\alpha]}=0.3$ (solid line), $Im{[\alpha]}=0.7$ (dashed line), and  $Im{[\alpha]}=1.5$ (short-dashed line). (a)  $\lambda_{1}/|\lambda_{2}|=0$. Curve points bellow $1$ clearly signal entanglement of the circularly polarized modes (b) $\lambda_{1}/|\lambda_{2}|=0.1$. The criterion is affected by the non-Hermiticity of the Hamiltonian, but at initial times it is still robust signaling entanglement, as depicted in the inset.}
\label{fig5}
\end{figure}

\begin{figure}[!ht]
\includegraphics[scale=0.4]{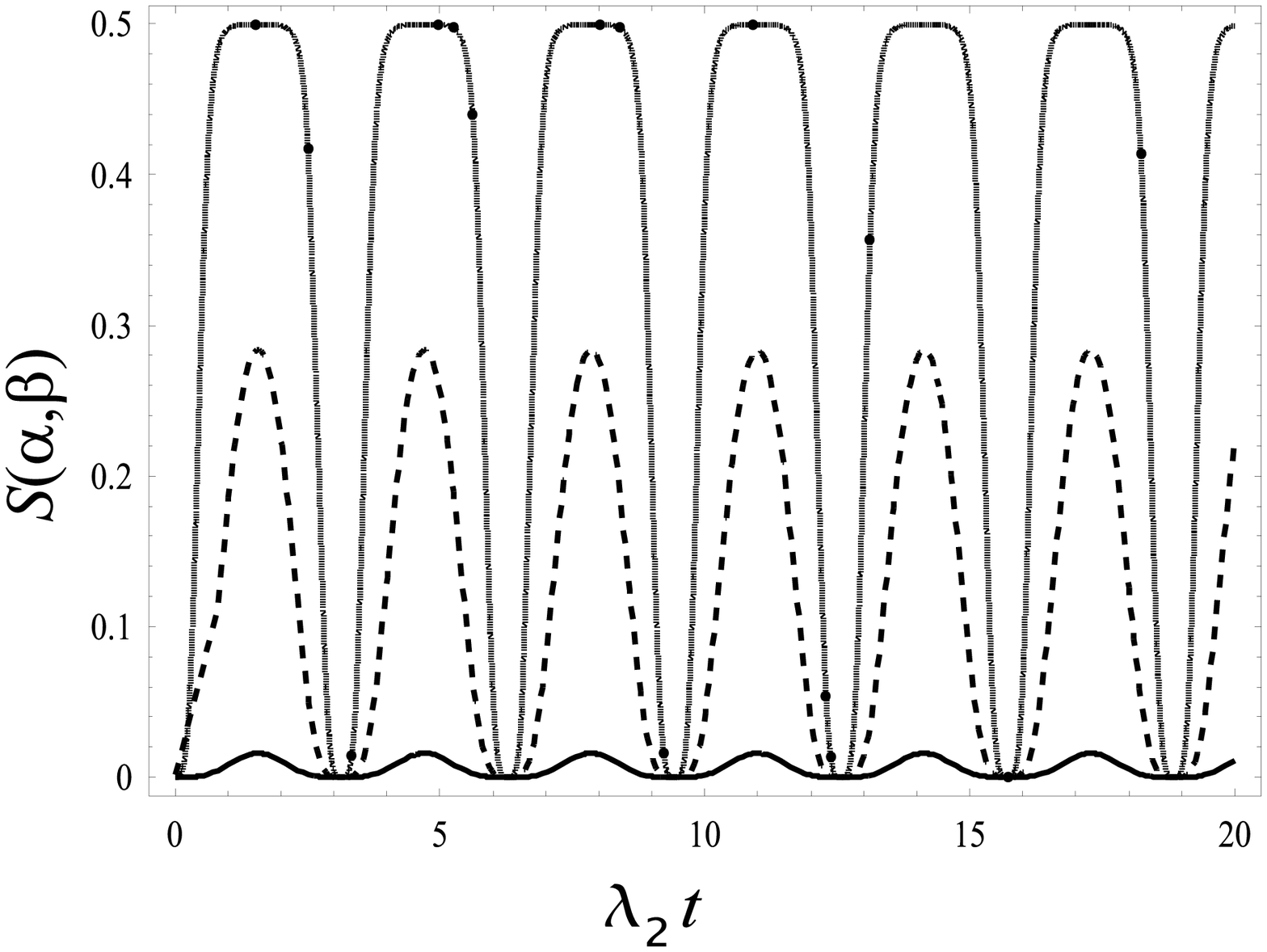}

\vspace{-4cm}
\includegraphics[scale=0.4]{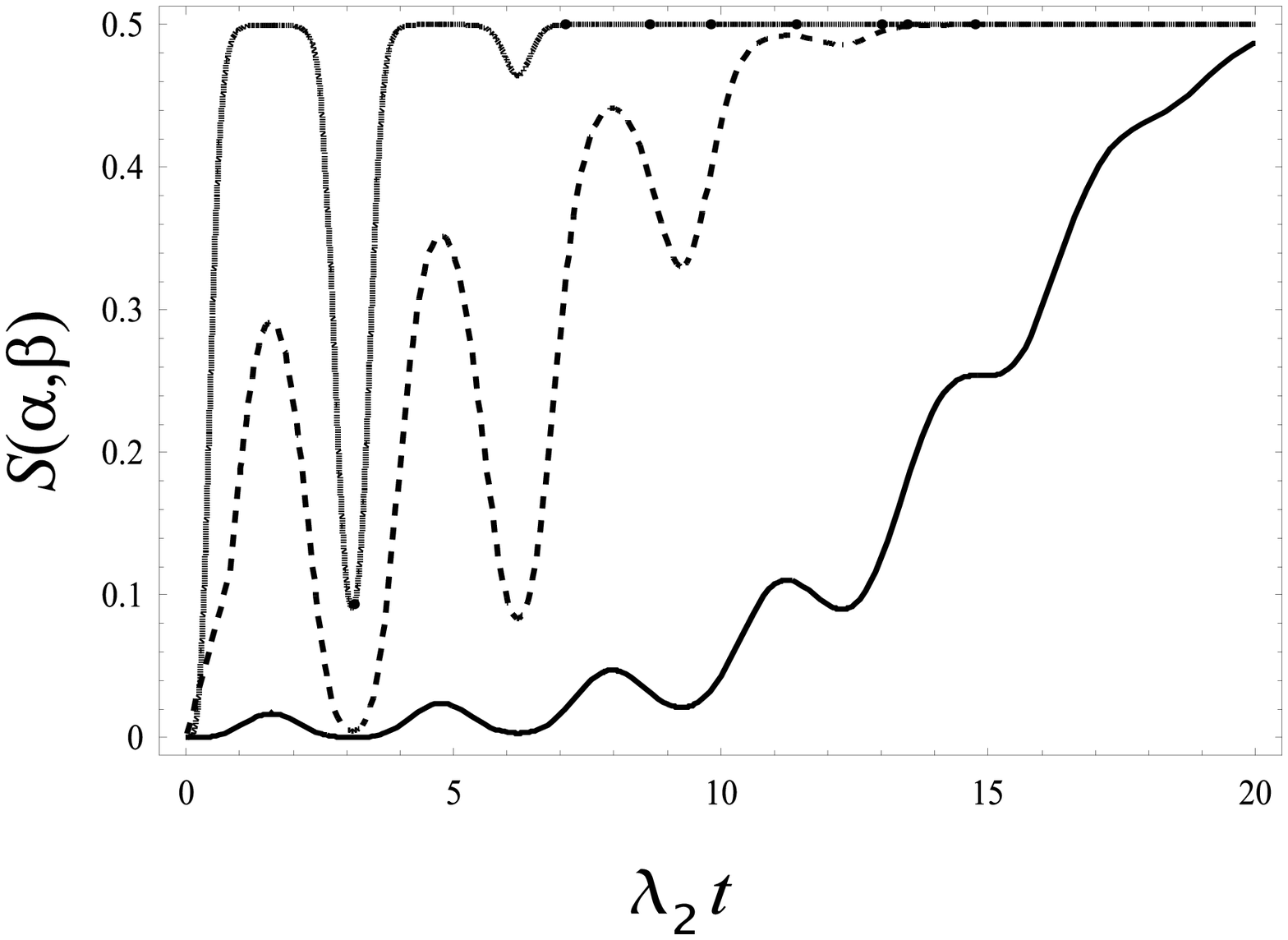}\vspace{-4cm}
\caption{Plot of linear entropy {time evolution} for the + circular polarization mode for $\alpha=\beta$ fixing three different lines for $Re{[\alpha]}=0$, $Im{[\alpha]}=0.3$ (continuous line), $Im{[\alpha]}=0.7$ (dashed line), and  $Im{[\alpha]}=1.5$ (short-dashed line). (a)  $\lambda_{1}/|\lambda_{2}|=0$ Indicates a periodic entanglement and disentanglement of the two orthogonal modes. (b) $\lambda_{1}/|\lambda_{2}|=0.1$ The periodicity of entanglement-disentanglement is affected by the non-Hermiticity of the Hamiltonian. The system tends to be highly entangled as the time goes on.}
\label{fig6}
\end{figure}

Since the $x$ polarized mode is always in a vacuum state the variances of its quadratures
will not change with time. Thus the fact that one of the variances of one of the modes is
squeezed bellow the noise limit is a good indicator that entanglement may be occurring in
the circular polarization. Indeed if we plot the inseparability criterion for the circular
polarization
\br
I_{+,-}(\alpha,\beta)&=&\frac{1}{2}[\Delta^{2}(X_{+}+X_{-})(\alpha,\beta)\nn\\&&+\Delta^{2}(Y_{+}-Y_{-})(\alpha,\beta)],
\er
where the variances are obtained using the previous expressions for this polarization, we
observe a clear signature of entanglement, as can be seen in Fig.\ref{fig5}.
 We remark however that there are situations (parameter choices) where the criteria will not indicate entanglement, while the state is clearly entangled. This is not a surprising fact since the state is non-Gaussian and thus violation of criteria is not a necessary ingredient for the existence of entanglement.
For comparison we plot in Fig.\ref{fig6} the one mode reduced linear entropy,
$S(\alpha,\beta)=1-Tr\{\rho_+^2\}$ {time evolution}, which in this case, since the joint
system is pure, indicates entanglement between the modes. It is clear to notice that the
criterion is correctly indicating entanglement for the same situations considered in
Fig. \ref{fig5}.

\begin{figure}[!ht]
\includegraphics[scale=0.3]{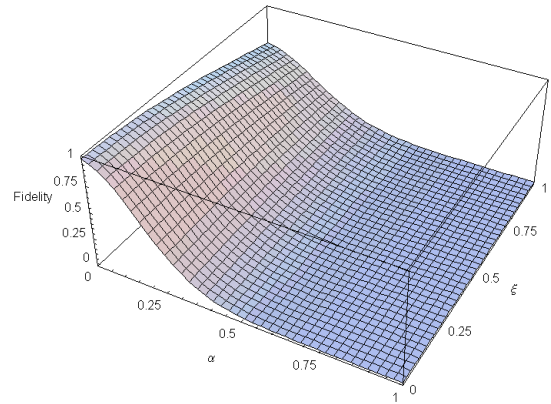}
\caption{\footnotesize (Color online) 3D plot of the fidelity between the generated state (Eq.\ref{plin4}) and the squeezed vacuum state (Eq.\ref{vac}) {as a function of $\alpha$ and $\xi$}. The coupling condition adopted for the plot is
$\lambda_{1}/|\lambda_{2}|\approx 0$.}
\label{fig7}
\end{figure}

It is interesting to remark that the points of Figs. \ref{fig5} and
\ref{fig6} that indicate maximal entanglement are given by the state
(\ref{ecs111}) for the circular polarization, which on its turn
corresponds to a superposition of small amplitude coherent state
(\ref{plin4}) for the linear polarization. As we discussed this
state has presented squeezing of one of its quadratures variance.
Now we want to question how far is this state from the vacuum
squeezed state, commonly attributed as being the case in many
experimental situations? As a matter of fact, by following the
discussion in \cite{GR06}, the fidelity, $
F(|\xi\rg,\rho_{cat})=\sqrt{\lg\xi |\rho_{cat}|\xi\rg}
$, between this superposition
state in the $y$ polarization and the squeezed vacuum state,
\br
|\xi\rg=\frac{1}{\xi}\sum_{n=0}^{\infty}\lbk\frac{\sqrt{(2n)!}}{n!}\lbk -\frac{1}{2}e^{i\theta}\tanh{(|\xi|)}\rbk^{n}\rbk |2n\rg,
\label{vac}
\er
can be very close to $1$ as depicted in Fig. \ref{fig7}, and in some sense they
are ``similar'' states.
%
For small values of  $\alpha$  (which represents the states generated in the
mentioned experiment), the fidelity (with the squeezed vacuum state with small squeezing
parameter) is very high. However, it is important to notice that the superposition state
is {\sl not} a Gaussian state, although the squeezed vacuum state {\sl is} Gaussian.
The fact of being or not Gaussian is important for the criterion of entanglement used to
interpret the results in Josse {\sl et al} \cite{Josse1}. Moreover the Entanglement of
Formation \cite{Gie03} expression for symmetric Gaussian states, employed to quantify
entanglement in Ref. \cite{Josse1}, in this case is only a lower bound for the entanglement
of the two modes.

\section{Conclusion}

In this paper we have investigated a model \cite{Josse0} for the
interaction of two quantum fields with {an ensemble of}
$X$-like four level {atoms}, and we derived an effective
Hamiltonian accounting for the field modes interaction. We have
demonstrated that the Heisenberg equations for the field modes
operators are non-linear leading to a non-Gaussian evolution. This
non-linear evolution will lead to an entangled non-Gaussian state or
to a non-Gaussian superposition of coherent states, when viewed from
the circular or linear polarization reference frame, depending on
the initial states for the two modes and atomic system. Even when
the evolution is kept as close as possible from a Gaussian one,
i.e., bilinear in the two field modes operators, a superposition
state of two atomic degenerate fundamental collective
states can lead to a non-Gaussian evolution. By appropriately
setting one of the input linearly polarized modes in the vacuum
state, we obtain in the output a coherent superposition of coherent
states in the orthogonally polarized linear modes. Although this
state is non-Gaussian it preserves similarities with the one-mode
squeezed vacuum state. When viewed from the circular polarization
frame, this superposition results in an entangled coherent state
between the two modes in polarization $+$ and $-$. We have { 
compared qualitatively } these results to recent experimental results 
\cite{Josse1} with a similar system, which however have attributed a 
Gaussian nature to the Quantum fields. The presence of this superposition, 
in one polarization reference frame and an entangled state in the other can
explain all the non-classical features observed in this experiment,
with a non-Gaussian state however.

Linearization procedures for field operators are common in quantum
optics whenever non-linear processes are present. While it can lead
to a good approximation to non-classical features such as squeezing,
it does not allow to correctly infer the non-Gaussian nature of the
system. By linearizing one is at most mimicking the second order
moments which are present in the most general nonlinear case. Thus
it is commonly attributed a Gaussian evolution to the system as
well, which as we have showed it is not particularly correct.
Inference on quantum properties of single systems may well be
described by the linearization procedure, but the description of
entanglement of (at least) two modes suffers from inconsistencies
which are only resolved when dealing with the correct nonlinearized
case. One typical feature that cannot be correctly inferred is the
separability of the two modes, since the criterion employed is only
necessary and sufficient for Gaussian states. We have shown that
invariably in the present system the quantum state is non Gaussian.
Remarkably recently much effort has been dedicated to the
generation of such non-Gaussian states in propagating light fields
by photon-subtraction \cite{GR06,POL06}. As we demonstrated, in
principle { such type of states} may  be generated in the
experiments reported in \cite{Josse3,Josse0,Josse1,Josse2} and any
other similar situation, whenever the atoms can be prepared in coherent 
superpositions if nonlinearities are to be neglected. If the nonlinearities 
are present, the entangled two mode field state will always be non-Gaussian 
independently of the atomic system state preparation.

\acknowledgments{RJM acknowledges financial support from CAPES. KF
and MCO are partially supported by CNPq.}


\begin{thebibliography}{99}
\bibitem{BP03} S.L. Braunstein and A.K. Pati,
\textit{Quantum Information with Continuous Variables}
(Kluwer Academic, Dordrecht, 2003).

\bibitem{BL05} S.L. Braunstein  \and P. van Loock,
Rev. Mod. Phys. \textbf{77}, 513 (2005).

\bibitem{BK98} S.L. Braunstein and H. J. Kimble,
Phys. Rev. Lett. \textbf{80}, 869 (1998).

\bibitem{FU98} A. Furusawa, J. L. Sørensen, S. L. Braunstein,
C. A. Fuchs, H. J. Kimble, and E. S. Polzik,
Science 282, 706 (1998).

\bibitem{BAN99} M. Ban,
J. Opt. B: Quantum Semiclass. Opt. \textbf{1}, L9 (1999).

\bibitem{BK00} S.L. Braunstein and H. J. Kimble,
Phys. Rev. A \textbf{61} 042302 (2000).

\bibitem{ZP00} J. Zhang and K. Peng,
Phys. Rev. A \textbf{62}, 064302 (2000).

\bibitem{Josse3} V. Josse, A. Dantan, L. Vernac, A. Bramati, M. Pinard,
and E. Giacobino,
Phys. Rev. Lett. {\bf 91}, 103601 (2003).

\bibitem{duan} L.-M. Duan, G. Giedke, J.I. Cirac, and P. Zoller,
Phys. Rev. Lett. {\bf 84}, 2722 (2000).
\bibitem{simon} R. Simon, Phys. Rev. Lett. \textbf{84}, 2726 (2000).

\bibitem{englert} B. G. Englert and K. W\'{o}dkiewicz,
Int. J. Quant. Inf. \textbf{1}, 153 (2003).

\bibitem{Josse0} V. Josse, A. Dantan, A. Bramati, M. Pinard, and
E. Giacobino,
J. Opt. B: Quantum Semiclass. Opt. {\bf 5}, S513 (2003).

\bibitem{Josse1} V. Josse, A. Dantan, A. Bramati, M. Pinard, and
E. Giacobino,
Phys. Rev. Lett. {\bf 92}, 123601 (2004).

\bibitem{Josse2} V. Josse, A. Dantan, A. Bramati, and E. Giacobino,
J. Opt. B: Quantum Semiclass. Opt. {\bf 6}, S532 (2004).


\bibitem{lezama} A. Lezama, P. Valente, H. Failache, M. Martinelli, and P. Nussenzveig, Phys. Rev. A \textbf{77}, 013806 (2008).

\bibitem{matsko} A. B. Matsko, I. Novikova, G. R. Welch, D. Budker,
D. F. Kimball, and S. M. Rochester,
Phys. Rev. A ${\mathbf{ 66}}$, $043815$ ($2002$).

\bibitem{GR06} A. Ourjoumtsev, R. Tualle-Brouri, J. Laurat, and
P. Grangier,
Science {\bf 312}, 83 (2006).

\bibitem{POL06} J. S. Neergaard-Nielsen, B. M. Nielsen, C. Hettich,
K. M\o lmer, and E. S. Polzik,
Phys. Rev. Lett. \textbf{97}, 083604 (2006).

\bibitem{TUR95} Q. A. Turchette, C. J. Hood, W. Lange, H. Mabuchi,
and H.J. Kimble,
Phys. Rev. Lett. {\bf 75}, 4710 (1995).

\bibitem{agarwal} G. S. Agarwal, P. Lougovski, H. Walther,
J. Mod. Opt. ${\mathbf{ 52}}$, $1397$ ($2005$).

\bibitem{entmilburn} M. C. de Oliveira and G.J. Milburn, Phys. Rev. A {\bf 65}, 032304 (2002).


\bibitem{kuz1}
A. Kuzmich, K. M\"olmer, and E. S. Polzik, Phys. Rev. Lett. \textbf{79}, 481 (1998).

\bibitem{kuz2} A. Kuzmich, N. P. Bigelow, and L. Mandel, Europhys. Lett. A \textbf{42}, 481 (1998).

\bibitem{lukin1} M. D. Lukin, S. F. Yelin, and M. Fleischhauer, Phys. Rev. lett. \textbf{84}, 4232 (2000).
\bibitem{duan2} L. M. Duan, J. I. Cirac, P. Zoller, and E. S. Polzik, Phys. Rev. Lett. \textbf{85}, 5643 (2000).
\bibitem{lukin2} L. M. Duan, M. D. Lukin, J. I. Cirac, and P. Zoller, Nature \textbf{414}, 413 (2001).
\bibitem{hald} J. Hald, J. L. Sorensen,C. Schori, and E. S. Polzik, Phys. Rev. Lett. \textbf{83}, 1319 (1999).


\bibitem{liu} C. Liu, Z. Dutton, C. H. Behroozi, and L. V. Hau, Nature \textbf{409}, 490 (2001).


\bibitem{bergmann} F. Vewinger, M. Heinz, R. G. Fernandez, N. V. Vitanov, and K. Bergmann, Phys. Rev. Lett. \textbf{91}, 213001 (2003).
\bibitem{knight} M. S. Kim, W. Son, V. Bu\v{z}ek, and P. L. Knight,
Phys. Rev. A ${\mathbf{ 65}}$, $032323$ ($2002$).

\bibitem{bin} W. Xiang-bin, Phys. Rev. A {\bf 66}, 024303 (2002).

\bibitem{Marcos1noncl} M. C. de Oliveira and W. J. Munro,
Phys. Lett. A \textbf{320}, 352 (2004).

\bibitem{dodonov} V. V. Dodonov, I. A. Malkin, and V.I. Man'ko,
 Physica {\bf 72}, 597 (1974).

\bibitem{marcosmunro}M. C. de Oliveira and W. J. Munro,
Phys. Rev. A \textbf{61}, 042309 (2000).

\bibitem{marcos1} M. C. de Oliveira,
Phys. Rev. A \textbf{72}, 012317 (2005).

\bibitem{sanders} B.C. Sanders, Phys. Rev. A {\bf 45}, 6811 (1992).
\bibitem{Gie03} G. Giedke, M. M. Wolf, O. Kr\"uger, R. F. Werner, and J. I. Cirac, Phys. Rev. Lett. \textbf{91}, 107901 (2003).
\end{thebibliography}
\end{document}